\documentclass[10pt]{elsart}
\usepackage[latin1]{inputenc}
\usepackage{amsmath}
\usepackage{amssymb}
\usepackage{amscd}
\usepackage{graphicx}
\usepackage{subfigure}

\def\sump{\sideset{}{'}\sum}

\def\O{{\mathcal O}}

\def\Z{\mathbb{Z}}
\def\R{\mathbb{R}}

\def\S{{\mathcal S}}
\def\F{{\mathcal F}}
\def\sign{\operatorname{sign}}
\def\KN{\operatorname{K}_0}
\def\KO{\operatorname{K}_1}

\def\Re{\operatorname{Re}}
\def\Im{\operatorname{Im}}

\def\nkl{n_{kl}}
\def\bN{\beta\rightarrow 0}
\def\ObN{\underset{\beta\rightarrow 0}{\mathcal O}}
\def\ddh{\text{2d\negthinspace+\negthinspace h}}

\newcommand\pg[1]{\psi^{(#1)}}
\def\lx{\lambda_x}
\def\ly{\lambda_y}
\def\lz{\lambda_z}
\def\mpt{\,\text{.}}
\def\mcm{\,\text{,}}

\theoremstyle{plain}
\newtheorem{prog}{Program}[section]
\newlength\picturewidth
\newlength\pictureheight
\newcommand\algo[1]{{\bf #1}}

\begin{document}

%%%%%%%%%%%%%%%%%%%%%%%%%%%%%%%%%%%%%%%%%%%%%%%%%%%%%%%%%%%%%%%%%%%%%%%%%%%%%%
%                    TITLE PAGE
%%%%%%%%%%%%%%%%%%%%%%%%%%%%%%%%%%%%%%%%%%%%%%%%%%%%%%%%%%%%%%%%%%%%%%%%%%%%%%
\begin{frontmatter}
 
  \title{MMM2D: A fast and accurate summation method for electrostatic
    interactions in 2D slab geometries}

  \author{Axel Arnold}\ead{arnolda@mpip-mainz.mpg.de},
  \author{Christian Holm}\ead{holm@mpip-mainz.mpg.de}

  \address{Max-Planck-Institut f\"ur Polymerforschung,
    Ackermannweg 10, D-55128 Mainz, Germany}
 
  \begin{abstract}
    We present a new method, in the following called \algo{MMM2D}, to
    accurately calculate the electrostatic energy and forces on charges
    being distributed in a two dimensional periodic array of finite
    thickness. It is not based on an Ewald summation method and as such
    does not require any fine-tuning of an Ewald parameter for
    convergence. We transform the Coulomb sum via a convergence factor
    into a series of fast decaying functions which can be easily
    evaluated.  Rigorous error bounds for the energies and the forces
    are derived and numerically verified. Already for small systems our
    method is much faster than the traditional $2D$--Ewald methods, but
    for large systems it is clearly superior because its time demand
    scales like $\O(N^{5/3})$ with the number $N$ of charges considered.
    Moreover it shows a rapid convergence, is very precise and easy to
    handle.
  \end{abstract}

\end{frontmatter}

%%%%%%%%%%%%%%%%%%%%%%%%%%%%%%%%%%%%%%%%%%%%%%%%%%%%%%%%%%%%%%%%%%%%%%%%%%%%%%
%                 Beginn main text
%%%%%%%%%%%%%%%%%%%%%%%%%%%%%%%%%%%%%%%%%%%%%%%%%%%%%%%%%%%%%%%%%%%%%%%%%%%%%%            
%%%%%%%%%%%% chapters

\section{Introduction}
The calculation of Coulomb interactions are time consuming due to their long
range nature. In principle, each charge $q_i$ at
position $p_i$ interacts with all others, leading to a computational
effort of $\O(N^2)$ already within the central simulation box.  For many
physical investigations one wants to simulate bulk properties and therefore
introduces periodic boundary conditions to avoid surface effects. The Coulomb
energy $E$ then has to be computed as a sum over all periodic images,
\begin{eqnarray}
  E = \frac{1}{2} \sump_{ n\in\Z^3} \sum_{i,j}
  \frac{q_i q_j}{|p_i - p_j + n \lambda|},\label{eq:coulomb}
\end{eqnarray}
where for simplicity we assume a cubic simulation box of length $\lambda$. The
prime denotes that for $n = 0$ the term $i=j$ has to be omitted. This sum is
merely conditionally convergent, meaning that its value depends on the
summation order. Normally a spherical limit is used, i.\ e. the vectors $n$
are added in the order of increasing length. Then $E$ can be computed for
example via the traditional Ewald summation method \cite{ewald21a}. The basic
idea of this method is to split the original sum via a simple transformation
involving the splitting parameter $\alpha$ into two exponentially convergent
parts, where the first one is short ranged and evaluated in real space, the
other one is long ranged and can be analytically Fourier transformed and
evaluated in Fourier space.  For any choice of the Ewald parameter $\alpha$
and no truncation in the sums the formula yields the exact result.  In
practice one cuts off the infinite sums at some finite values and obtains $E$
to a user controlled accuracy, which is possible by using error estimates for
the cut-offs \cite{kolafa92a}.

Another way of computing
Eq.~\eqref{eq:coulomb} is via a convergence factor
\begin{eqnarray}
  \label{eq:convergence}
  \tilde E = \lim_{\beta \rightarrow 0 } \frac{1}{2} \sump_{n\in\Z^3} \sum_{ij}
  \frac{q_i q_j \exp{ (-\beta |p_i - p_j + n \lambda|) }}
  {|p_i - p_j + n \lambda|}.
\end{eqnarray}
This approach is used in a method due to Sperb \cite{sperb98a} called
\algo{MMM}.

In Refs.\cite{deleeuw80a,deleeuw80b} it was shown that Eqs.~\eqref{eq:coulomb}
and \eqref{eq:convergence} differ by a term proportional to the dipole moment
for 3D systems. Sperb et al. have developed a factorization approach which
yields an $\O(N \log N)$ algorithm \cite{sperb01a}, comparable in speed to the
FFT mesh based Ewald algorithms\cite{hockney88a,deserno98a}. An different
convergence factor was used by Lekner \cite{lekner91a} to efficiently sum up
the 3D Coulomb sum. But this approach still leads to an $\O(N^2)$ algorithm.

For thin polyelectrolytes films, membrane interactions, or generally confined
charged systems, one is interested in summations where only 2 dimensions
are periodically replicated and the third one (here always the z-direction) is
of finite thickness $h$ ($2D + h$ geometry). For this geometry Ewald based
formulas are only slowly convergent, have mostly $\O(N^2)$ scalings and
no ``a priori'' error estimates exist \cite{widmann97a}. In
Ref.\cite{widmann97a} a comparison was made with respect to speed and accuracy
of the three most popular versions, the traditional $2d$--Ewald method
developed by Heyes, Barber, and Clarke (HBC) \cite{heyes77a}, the one due to
Hautmann-Klein (HK) \cite{hautman92a}, and the approach attributed to Nijboer
and de Wette (NdW) \cite{nijboer57a,nijboer84a}.  Another variant is due to
Smith \cite{smith88a}, which combines HBC and NdW, resulting in a good
scaling, but still slow convergence. Recently two more Ewald-type methods were
proposed with different ways of approximating the Fourier space sum of the HBC-formula\cite{kawata01a,kawata01b}.

The aim of this paper is to develop a new method, which we call \algo{MMM2D}.
A brief description of the key features of this algorithm was already given in
Ref.\cite{arnold01c}.  \algo{MMM2D} will be using the same ideas leading to
\algo{MMM} and is therefore derived from an equation like
Eq.~\eqref{eq:convergence} using a convergence factor. We will show that for
two dimensional systems this approach yields the same results as the spherical
limit, i.\ e. $E=\tilde E$.  In principle, already the 3D formula developed by
Lekner \cite{lekner89a} could be used for 2D + h Systems.  It employs modified
Bessel functions, but is only useful for particles separated in the
$z$--direction, and is only calculable pairwise. Sperb \cite{sperb94a} derived
a different formula, which is also applicable for particles lying in the same
$z$--plane. A combination of theses two formula has been used recently in MD
simulations, because it is easy to implement \cite{csajka00a} for small number
of particles, but suffers the drawback of being an $\O(N^2)$ method, and
therefore the system sizes which can be investigated are limited. Moreover
these methods use different convergence factors, and up to now no
investigations have been performed to see whether both methods lead to the
same result as the one obtained by using a spherical limit.

To precisely define the problem we consider systems of $N$ particles with
charges $q_i\in\R$ and pairwise different coordinates $p_i=(x_i,y_i,z_i)^T \in
B_0$, $i=1,\hdots,N$, where
\begin{equation*}
  B_0=
  \biggl(-\frac{\lx}{2},\frac{\lx}{2}\biggr]\times
  \biggl(-\frac{\ly}{2},\frac{\ly}{2}\biggr]\times\R\subset\R^3
\end{equation*}
is the simulation box. Furthermore it is assumed that the system
is charge neutral, i.\ e. $\sum_{i=1}^N q_i = 0$.

Now let $\nkl:=(k\lx, l\ly, 0)^T$ for $k,l\in\Z$. The expressions to
be calculated are the interaction energies of particle $i$ with all other
charges, defined as
\begin{equation}
  \label{ordered2d}
  E_i=\sum_{S=0}^\infty
  \sum_{\substack{k,l\in\Z,\\k^2+l^2=S}}
  \sump_{j=1}^N \frac{q_i q_j}{|p_i - p_j + \nkl|}\quad
  i=1\hdots N\mpt
\end{equation}
As usual the prime $'$ on the inner sum indicates that the term $j=i$ for
$k=l=0$ is omitted.  

In Sec.~\ref{sec:algo} we will develop the formulas used by
\algo{MMM2D} for calculating the energy and the forces. In
Sec.~\ref{sec:errors} we develop error formulas for the energy and
force calculations, which can be easily calculated prior to a
simulation.  Then, in Sec.~\ref{sec:implementation} we give the
details of our implementation, which is based on a factorization
approach resulting in an $\O(N^{5/3})$ scaling.  In
Sec.~\ref{sec:numerics} we apply our algorithm to several test cases
and prove its applicability and efficiency.  We will also demonstrate
that it is much faster than the $2d$--Ewald method, and is thus a
superior alternative to all aforementioned methods. We will end with
some conclusions, and show for the interested reader in the Appendix A
that for two dimensional systems $E=\tilde E$, i.\ e. that the
summation using a convergence factor as in Eq. \eqref{eq:convergence}
yields exactly the same result as the summation using a spherical
limit as the $2d$--Ewald method, there is no dipolar correction term needed.

\section{The \algo{MMM2D} Method}\label{sec:algo}
The transformations given here are similar to the
transformations used in R. Strebel's \algo{MMM}
\cite{strebel99a,sperb01a} for three--dimensional periodic systems.
The ideas used in the following are similar to the ideas used to
develop the formulas for \algo{MMM}. Basically, two different formulas are
used to calculate the energy and forces. For the interaction between
particles that are well separated, a formula that can be calculated
in linear time is used. For the interaction between neighboring
particles a second formula is used. This leads to an $\O(N^{7/5})$
algorithm. Up to here the same ideas are used for \algo{MMM2D}. But for the
three dimensional case some further tricks are used to achieve the
scaling of $\O(N\log N)$. These tricks involve the symmetry of the
coordinates which is not true for two dimensional systems. Therefore
this better scaling cannot be carried over to the two dimensional case.

The formulas needed for \algo{MMM2D} are derived in the next subsections in
the same way as has been done for \algo{MMM}\cite{strebel99a}.  We first
rewrite the sum \eqref{ordered2d} using a convergence factor:
\begin{equation}
  \label{damped2d}
  E_i=\,\lim_{\bN}\sum_{k,l\in\Z}\sump_{j=1}^N \frac{q_i
    q_je^{-\beta|p_{ij} + \nkl|}}
  {|p_{ij} + \nkl|}
  =\,\lim_{\bN}\sum_{j=1}^N q_iq_j\phi_\beta(x_{ij},
  y_{ij},z_{ij})
\end{equation}
where $p_{ij}=(x_{ij},y_{ij},z_{ij})=p_i-p_j$,
\begin{equation}
  \label{pwdamped2d}
  \phi_\beta(x,y,z)=\,\tilde\phi_\beta(x,y,z) +
  \begin{cases}
    \frac{e^{-\beta r}}{r}
    & (x,y,z) \neq (0,0,0) \\
    0 & (x,y,z) = (0,0,0)
  \end{cases}
\end{equation}
and
\begin{equation}
  \label{pwdamped2dsm}
  \begin{split}
    \tilde\phi_\beta(x,y,z)=\,&\sum_{(k,l)\neq (0,0)}
    \frac{e^{-\beta r_{kl}}}
    {r_{kl}}
  \end{split}\mcm
\end{equation}
where we abbreviated
\begin{gather*}
  r_{kl} = \sqrt{(x + k \lx)^2 + (y + l \ly)^2 + z^2}\mcm\quad
  r_k    = r_{k0}\mcm\quad r = r_0\mpt
\end{gather*}
The fact that the rhs. of Eq. \eqref{damped2d} is equal to Eq.
\eqref{ordered2d} is non--trivial; see appendix~\ref{prfconv} for a proof.

\subsection{Transformation of $\phi_\beta$ for $z\neq 0$}
First we concentrate on developing an absolutely and rapidly converging
formula for $\phi_\beta$. Then we can easily form the
limit $\bN$ and obtain a formula for $\phi$.
For $z\neq 0$ and $\beta>0$ the sum in $\phi_\beta$ is an absolutely
convergent sum of Schwartz class functions. Therefore for $\delta > 0$ and $x
\in \R$, we
can apply the \emph{Poisson formula}:
\begin{equation}
  \label{poisson}
  \sum_{k\in\Z}f(x + \delta k)=
  \frac{1}{|\delta|}\sum_{p\in\Z}\F(f)
  \left(\frac{p}{\delta}\right)
  e^{2\pi i \frac{p}{\delta} x}\mcm
\end{equation}
where $\F$ denotes the Fourier transformation. Furthermore we will be using
the formulas
\begin{align*}
  \F\left(\frac{e^{-\beta\sqrt{\alpha^2+\cdot\,^2}}}
    {\sqrt{\alpha^2+\cdot\,^2}}\right)
  &= 2\KN\left(\alpha\sqrt{\beta^2+(2\pi\,\cdot)^2}\right)\mcm \\
  \F\left(\KN\left(\alpha\sqrt{z^2+\cdot\,^2}\right)\right)
  &= \pi\frac{e^{-z\sqrt{\alpha^2+(2\pi\,\cdot)^2}}}
  {\sqrt{\alpha^2+(2\pi\,\cdot)^2}}
\end{align*}
which are valid for $\alpha,z\in\R$ and can be found, for
example, in \cite{oberhett}. $\KN$ is called the modified Bessel
function of order 0. For properties of the Bessel functions, see
\cite{abramowitz}.

Setting
\begin{gather*}
  \beta_{pq}=\sqrt{\beta^2+(2\pi u_x p)^2+(2\pi u_y q)^2}\mcm\\
  \beta_p   =\sqrt{\beta^2+(2\pi u_x p)^2}\quad\text{and}\quad
  \beta_q   =\sqrt{\beta^2+(2\pi u_y q)^2}\mcm\\
  \rho_l = \sqrt{(y + l \ly)^2 + z^2}\mcm\quad \rho = \rho_0\mcm\quad
  u_x    = \frac{1}{\lx} \quad\text{and}\quad u_y    = \frac{1}{\ly}\mpt
\end{gather*}
we obtain after two Fourier transformations
\begin{equation*}
  \begin{split}
    \phi_\beta(x,y,z)
    &=\,  \sum_{k,l\in\Z} \frac{e^{-\beta r_{kl}}}{r_{kl}}
    =\,\sum_{l\in\Z}\left(\sum_{k\in\Z}
      \frac{e^{-\beta r_{kl}}}{r_{kl}}\right)\\
    &=\,  2 u_x \sum_{p\in\Z}
    \left(
      \sum_{l\in\Z}\KN\left(\beta_p \rho_l \right)
    \right) e^{2\pi i u_x p x}\\
    &=\,  2\pi u_x u_y\sum_{p,q\in\Z}
    \frac{e^{- \beta_{pq}|z|}}
    {\beta_{pq}}
    e^{2\pi i u_x p x}e^{2\pi i u_y q y}\mpt
  \end{split}
\end{equation*}
Expanding the term for $p=q=0$ we find
$\frac{e^{-\beta|z|}}{\beta}= \beta^{-1} - |z| + \ObN(\beta)$, and
obtain our final formula
\begin{equation}
  \label{far}
  \begin{split}
    &\phi_\beta(x,y,z) =\, 8\pi u_x u_y\sum_{p,q>0}
    \frac{e^{-\beta_{pq}|z|}}{\beta_{pq}}
    \cos(2\pi u_x p x)\cos(2\pi u_y q y)\, +\\
    &\quad 4\pi u_x u_y\sum_{q>0}
    \frac{e^{-\beta_q|z|}}{\beta_q}
    \cos(2\pi u_y q y)\, +
    4\pi u_x u_y\sum_{p>0}
    \frac{e^{-\beta_p|z|}}{\beta_p}
    \cos(2\pi u_x p x)\, -\\
    &\quad 2\pi u_x u_y |z| + 2\pi u_x u_y \beta^{-1} + \ObN(\beta)\mpt
  \end{split}
\end{equation}

It has a singularity of $2\pi u_x u_y \beta^{-1}$ which is independent of the
particle coordinates. However, once the sum of $\phi_\beta$ is taken over all
particles, the singularity vanishes via the charge neutrality condition. For
the other parts of Eq.~\eqref{far} taking the limit $\bN$ is trivial. The
resulting formula (see Sec.~\ref{sec:engexpr}) can be evaluated linearly in
time, just like the Fourier part of the Ewald sum. This will be shown in more
detail in section~\ref{sec:implementation}.

Unfortunately the convergence becomes slower with decreasing $z$ and for
$z=0$ the sum is not defined. Thus we will need an alternative method
for small $z$.

\subsection{Transformation of $\tilde\phi_\beta$ for $z\approx 0$}
For small particle distances, the term for $k=l=0$ is dominant (and
must be omitted for the interaction of a particle with its own
images). Therefore we leave it out for now and  concentrate on
a rapidly convergent formula for $\tilde\phi_\beta(x,y,z)$.
This requires a little more work. For a more
detailed derivation see \cite{strebel99a,arnold01a}.

Since we omit the $k=l=0$ term, the area to sum over has a hole.
To efficiently treat the sum over this area we
split $\tilde{\phi}_\beta(x,y,z)=\Sigma_1+\Sigma_2$ where
\begin{equation*}
  \Sigma_1=\sum_{l\neq 0}\sum_{k\in\Z}\frac{e^{-\beta r_{kl}}}{r_{kl}}
  \quad\text{and}\quad
  \Sigma_2=\sum_{k\neq 0}\frac{e^{-\beta r_{k0}}}{r_{k0}}\mpt  
\end{equation*}

Graphically this can be displayed like that:
\begin{center}
  \includegraphics[height=3cm]{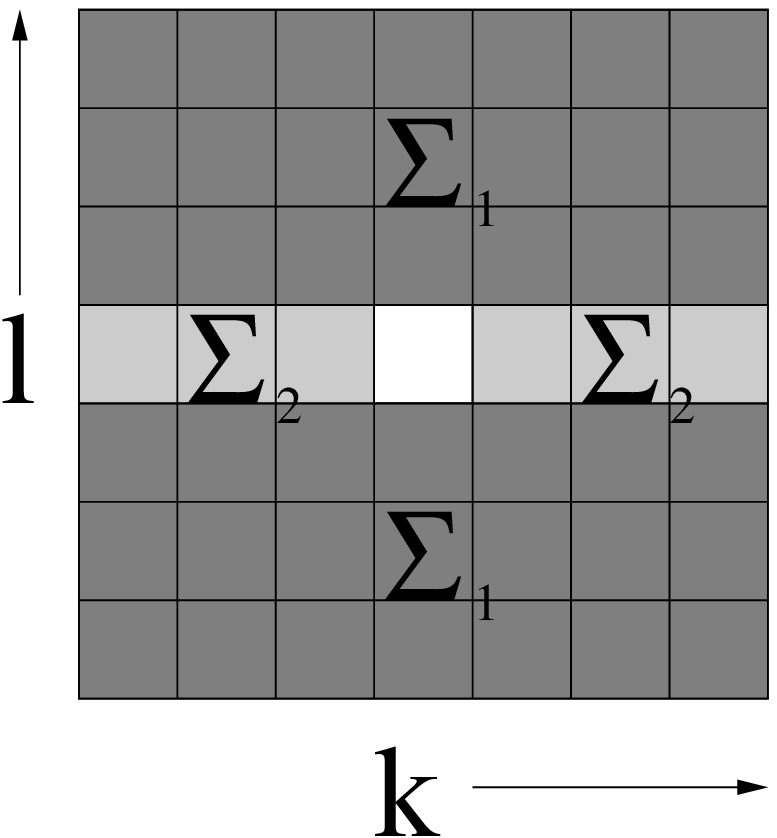}
\end{center}

We start by calculating $\Sigma_1$. Using the same argument as for formula
\eqref{far} we obtain
\begin{equation*}
  \begin{split}
    \Sigma_1
    &=\,  \sum_{l\neq 0}\sum_{k\in\Z}\frac{e^{-\beta r_{kl}}}{r_{kl}}
    =\,  2u_x\sum_{l\neq 0}\sum_{p\in\Z}\KN(\beta_p\rho_l)
    e^{2\pi i u_x p x}\\
    &=\,  2u_x\sum_{l,p\neq 0}\KN(\beta_p\rho_l)e^{2\pi i u_x p x} +
    2u_x\sum_{l\neq 0}\KN(\beta\rho_l)\mpt
  \end{split}
\end{equation*}

While the first sum converges fast, the second one is still
singular in $\beta$ and has to be investigated further:
\begin{equation*}
  \begin{split}
    2u_x\sum_{l\neq 0}\KN(\beta\rho_l)
    &=\,  2u_x\sum_{l\in\Z}\KN(\beta\rho_l) - 2u_x\KN(\beta\rho)\\
    &=\,  2u_x\sum_{l\in\Z}\KN(\beta\rho_l) - 2u_x(\log 2 - \gamma -
    \log(\beta\rho))+\ObN(\beta^2)
  \end{split}
\end{equation*}
where we have used the asymptotic behavior
$\KN(x)=-\log\frac{x}{2} - \gamma + \O(x^2)$ for $x\rightarrow 0$.

The first sum can now be Fourier transformed again:
\begin{equation*}
  \begin{split}
    \sum_{l\in\Z}\KN(\beta\rho_l)
    &=\,  \pi u_y\sum_{q>0}\frac{e^{-\beta_q |z|}}{\beta_q}
    \left(e^{2\pi i u_y q y}+e^{-2\pi i u_y q y}\right) +
    \pi u_y\frac{e^{-\beta|z|}}{\beta} \\
    &=\,2\pi u_y\Re\left(\sum_{q>0}\frac{e^{-2\pi u_y q |z|}}{2\pi u_y q}
      e^{2\pi i u_y q y}\right)+
    \pi u_y(\beta^{-1} - |z|) +\ObN(\beta)\mpt
  \end{split}
\end{equation*}

If we set $\zeta:=2\pi u_y(|z|+iy)$, we obtain
\begin{equation*}
  \begin{split}
    2\pi u_y\sum_{q>0}\frac{e^{-2\pi u_y q |z|}}{2\pi u_y q}
    e^{2\pi i u_y q y}
    &=\, \sum_{q>0}\frac{e^{-q\zeta}}{q} =\, -\log(\zeta) + \frac{\zeta}{2} -
    \sum_{n\ge 2}\frac{b_n}{n\,n!}\zeta^n
  \end{split}
\end{equation*}
where $b_n$ are the Bernoulli numbers. This series expansion is
valid only for $|\zeta|\le 2\pi$, which is fulfilled if
$|z|\le\frac{\ly}{2}$. The last equality is found by integration from
\begin{equation*}
  \begin{split}
    &\frac{d}{dz}\log\left(\frac{1-e^{-z}}{z}\right)=
    z^{-1}\left(\frac{z}{e^z - 1} - 1\right)\\
    &\quad= z^{-1}\left(\sum_{n=0}^\infty \frac{b_n}{n!}z^n - 1\right)
    = -\frac{1}{2} + \sum_{n=1}^\infty\frac{b_{2n}}{(2n)!}z^{2n-1}\mcm
  \end{split}
\end{equation*}
where the defining series $\frac{z}{e^z-1}=\sum_{n=0}^\infty
\frac{b_n}{n!}z^n$ for the Bernoulli numbers was used.

Using $\Re(-\log\zeta)=-\log|\zeta|=-\log(2\pi u_y \rho)$ and
$\Re\left(\frac{\zeta}{2}\right)=\pi u_y |z|$ we obtain
\begin{equation*}
  \begin{split}
    \sum_{l\in\Z}\KN(\beta\rho_l)=\,
    & -\sum_{n\ge 2}\frac{b_n}{n\,n!}\Re\left(\zeta^n\right)
    - \log(2\pi u_y \rho) + \pi u_y \beta^{-1} + \ObN(\beta)\mpt
  \end{split}
\end{equation*}

It is easy to see that $\zeta$ can be replaced by
$\xi:=2\pi u_y(z+iy)$ without changing the value of the sum.
This is of advantage for the calculation of the forces by
differentiation.

Combining everything we obtain
\begin{equation}
  \label{near_sigma1}
  \begin{split}
    \Sigma_1=\,
    &   2u_x\sum_{l,p\neq 0}\KN(\beta_p\rho_l)
    e^{2\pi i u_x p x} -
    2u_x\sum_{n\ge 2}\frac{b_n}{n\,n!}
    \Re\bigl((2\pi u_y (z + iy))^n\bigr) -\\
    &   2u_x\log(4\pi u_y) + 2u_x\log(\beta) + 2\pi u_x u_y\beta^{-1} +
    2u_x\gamma + \ObN(\beta)\mpt
  \end{split}
\end{equation}

Now we concentrate on $\Sigma_2 = \sum_{k > 0}\frac{e^{-\beta r_k}}{r_k} +
\sum_{k < 0}\frac{e^{-\beta r_k}}{r_k}$.
It is sufficient to investigate the first of the two sums, because
by replacing $x$ by $-x$ the value of the second sum can be obtained.

We start with
\begin{equation*}
  \begin{split}
    \sum_{k > 0}\frac{e^{-\beta r_k}}{r_k}
    &=\, \sum_{k > 0}e^{-\beta r_k}\left(\frac{1}{r_k} -
      \frac{1}{k\lx}\right) +
    \sum_{k > 0}\frac{e^{-\beta r_k}}{k\lx} \\
    &=\, \sum_{k > 0}\left(\frac{1}{r_k} -
      \frac{1}{k\lx}\right) + \ObN(\beta) +
    \sum_{k > 0}\frac{e^{-\beta r_k}}{k \lx}\mpt
  \end{split}
\end{equation*}
Details about the precise derivation of the equation can again be found in
\cite{strebel99a,arnold01a}.

Moreover by $r_k - k\lx=\underset{k\rightarrow\infty}{\O}(1)$ and
$\log\left(1-e^{-\alpha}\right)=\log\alpha + \O(\alpha)$ we obtain
\begin{equation*}
  \begin{split}
    \sum_{k > 0}\frac{e^{-\beta r_k}}{k\lx}
    =\, \sum_{k > 0}\frac{e^{-\beta k\lx}}{k\lx}
    \left(1+\ObN(\beta)\right)
    =\, -u_x\log(\lx\beta) + \ObN(\beta\log\beta))\mpt
  \end{split}
\end{equation*}

To evaluate the sum $\sum_{k > 0}\left(\frac{1}{r_k} -
  \frac{1}{k\lx}\right)$, we consider a $N_\psi$ such that
$N_\psi>u_x\rho + 1$. Then for $k \ge N_\psi$ we have
$\rho/|x+k\lx|\le 1$ and therefore we can use the binomial
series for $\sqrt{1 + \frac{\rho^2}{|x+k\lx|^2}}$ to obtain
\begin{equation*}
  \begin{split}
    &\sum_{k \ge N_\psi}\left(\frac{1}{r_k} - \frac{1}{k\lx}\right)\\
    &\quad=\, \sum_{k \ge N_\psi}\left(\frac{1}{|x+k\lx|} -
      \frac{1}{k\lx}\right) +
    \sum_{n>0}\binom{-\frac{1}{2}}{n}\sum_{k \ge N_\psi}
    \frac{\rho^{2n}}{|x+k\lx|^{2n+1}} \\
    &\quad=\, -u_x\pg{0}(N_\psi + u_x x) - u_x\gamma -
    u_x\sum_{n>0}\binom{-\frac{1}{2}}{n}
    \frac{\pg{2n}(N_\psi + u_x x)}{\lx^{2n}(2n)!}(u_x\rho)^{2n}
  \end{split}
\end{equation*}
where $\psi^{(n)}$ are the polygamma functions. For details on these
functions, see \cite{abramowitz}.

In summary we obtain for $\Sigma_2$:
\begin{equation*}
  \begin{split}
    \Sigma_2
    &=\, -u_x\sum_{n\ge 0}\binom{-\frac{1}{2}}{n}
    \frac{\left(\pg{2n}(N_\psi + u_x x) + 
        \pg{2n}(N_\psi - u_x x)\right)}{(2n)!}(u_x\rho)^{2n}-\\
    &\phantom{=\,} 2u_x\gamma + 2u_x\log(u_x) - 2u_x\log(\beta) +
    \sum_{k=1}^{N_\psi-1}\left(\frac{1}{r_k}+\frac{1}{r_{-k}}\right) +
    \ObN(\beta\log\beta))\mpt
  \end{split}
\end{equation*}

Combining the formulas for $\Sigma_1$ and $\Sigma_2$ we obtain,
for $|x|\le\frac{\lx}{2}$, $|y|,|z|\le \frac{\ly}{2}$ and
$N_\psi$ such that $N_\psi>u_x\rho + 1$,
the final formula
\begin{equation}
  \label{near}
  \begin{split}
    \tilde{\phi}_\beta(x,y,z)=\,
    &  4u_x\sum_{l,p>0}\left(\KN(\beta_p\rho_l) +
      \KN(\beta_p\rho_{-l})\right)\cos(2\pi u_x p x) -\\
    &  2u_x\sum_{n\ge 1}\frac{b_{2n}}{2n(2n)!}
    \Re\bigl((2\pi u_y (z+iy))^{2n}\bigr) +
    \sum_{k=1}^{N_\psi - 1}\left(\frac{1}{r_{k}} +
      \frac{1}{r_{-k}}\right) -\\
    &   u_x\sum_{n\ge 0}\binom{-\frac{1}{2}}{n}
    \frac{\left(\psi^{(2n)}(N_\psi + u_x x) +
        \psi^{(2n)}(N_\psi - u_x x)\right)}{(2n)!}(u_x\rho)^{2n} -\\
    &  2u_x\log(4\pi\frac{u_y}{u_x}) + 2\pi u_xu_y\beta^{-1}
    + \ObN(\beta\log\beta)\mpt
  \end{split}
\end{equation}
Of course this formula leads to the same singularity in $\beta$ as formula
\eqref{far} and the charge
neutrality argument also holds for any combination of the two formulas as
long as the sum is performed over all particles.

\subsection{Energy expressions}\label{sec:engexpr}
Now we want to give the full expressions for the energy after taking
the limit $\bN$. 

Setting
\begin{gather*}
  f_{pq} =\, \sqrt{(u_x p)^2 + (u_y q)^2}\mcm\quad f_p =\, u_x p
  \mcm\quad f_q =\, u_x q \mcm\\
  \omega_p=2\pi u_x p \quad\text{and}\quad \omega_q=2\pi u_y q 
\end{gather*}
we obtain for $z\neq 0$
\begin{multline}
  \label{fullfar}
  \phi(x,y,z) =\,
  4 u_x u_y\sum_{p,q>0}
  \frac{e^{-2\pi f_{pq}|z|}}
  {f_{pq}}
  \cos(\omega_p x)\cos(\omega_q y)\, +\\
  2 u_x u_y\left(\sum_{q>0}
    \frac{e^{-2\pi f_q|z|}}{f_q}
    \cos(\omega_q y)\, + \sum_{p>0}
    \frac{e^{-2\pi f_p|z|}}{f_p}
    \cos(\omega_p x)\right)\,
  - 2\pi u_x u_y |z|\mpt
\end{multline}
Eq. \eqref{fullfar} will be called the \emph{far formula}, since it
will be used only for particles that are separated in the $z$--plane
by a fixed minimum distance. See Sec.~\ref{sec:implementation} for
details of the implementation.

For $|z|\le\frac{\ly}{2}$ we have
\begin{equation}
  \label{fullnear}
  \begin{split}
    \tilde\phi(x,y,z)=\,
    & 4u_x\sum_{l,p>0}\left(\KN(\omega_p\rho_l) +
      \KN(\omega_p\rho_{-l})\right)\cos(\omega_p x) -\\
    &  2u_x\sum_{n\ge 1}\frac{b_{2n}}{2n(2n)!}
    \Re\bigl((2\pi u_y (z+iy))^{2n}\bigr)\,+\,
    \sum_{k=1}^{N_\psi-1}\left(\frac{1}{r_{k}} +
      \frac{1}{r_{-k}}\right) -\\
    &   u_x\sum_{n\ge 0}\binom{-\frac{1}{2}}{n}\frac{\left(
        \psi^{(2n)}(N_\psi + u_x x) +
        \psi^{(2n)}(N_\psi - u_x x)\right)}{(2n)!}(u_x\rho)^{2n} -\\
    &  2u_x\log\left(4\pi\frac{u_y}{u_x}\right)\mpt
  \end{split}
\end{equation}
In the same spirit as above this expression will be called the \emph{near
  formula}.

The far formula looks the same as an expression derived by Smith
\cite{smith88a} and which is attributed to an approach originally due to
Nijboer and de Wette \cite{nijboer57a}. Smith employed a spherical limit
instead of a convergence factor, and therefore obtained a totally different
singularity. Therefore his formula can be combined with any other formula
using a spherical limit for small $z$, for example $2D$--Ewald
\cite{heyes77a}. This combination leads then to the algorithm described in
detail in Ref.\cite{smith88a}.  Compared to this method, our combination has
the advantage that the near formula is faster to evaluate than the $2D$--Ewald
sum and possesses an easy to use error formula that will be derived in the
next section. Therefore the \algo{MMM2D} algorithm is faster than the Smith
algorithm, and in addition possesses full error control.

\subsection{The force}
Since the sums in equations \eqref{fullfar} rsp. \eqref{fullnear}
converge absolutely, the electrostatic force $F_i=-\nabla_{p_i}E_i$
can be derived by simple term-wise differentiation and the force
can be calculated as
\begin{equation*}
  F_i=\sum_{j=1}^Nq_iq_jF(x_i-x_j,y_i-y_j,z_i-z_j)
\end{equation*}
where for $z\neq 0\quad F=(F_x,F_y,F_z)^T$ is given by
\begin{equation}
  \label{forcesfar}
  \begin{split}
    F_x(x,y,z) =\,
    &   8\pi u_x^2 u_y\sum_{p,q>0} p
    \frac{e^{-2\pi f_{pq}|z|}}
    {f_{pq}}
    \sin(\omega_p x)\cos(\omega_q y)\, +\\
    &   4\pi u_x u_y\sum_{p>0}e^{-2\pi f_p |z|}
    \sin(\omega_p x)\mcm\\
    \intertext{\vfill} %%%%%%%%%%
    F_y(x,y,z) =\,
    &   8\pi u_x u_y^2\sum_{p,q>0} q
    \frac{e^{-2\pi f_{pq} |z|}}{f_{pq}}
    \cos(\omega_p x)\sin(\omega_q y)\, +\\
    &   4\pi u_x u_y\sum_{q>0}e^{-2\pi f_q|z|}
    \sin(\omega_q y)\mcm\\
    \intertext{\vfill} %%%%%%%%%%
    F_z(x,y,z) =\,
    &   8\pi\sign(z) u_x u_y\sum_{p,q>0}
    e^{-2\pi f_{pq}|z|}\cos(\omega_p x)\cos(\omega_q y)\, +\\
    &   4\pi\sign(z) u_x u_y\sum_{q>0} e^{-2\pi f_q|z|}
    \cos(\omega_q y)\, +\\
    &   4\pi\sign(z) u_x u_y\sum_{p>0} e^{-2\pi f_p|z|}
    \cos(\omega_p x)\, + 2\pi\sign(z) u_x u_y\mpt
  \end{split}
\end{equation}

For $|z|\le\frac{\ly}{2}$ we have
\begin{equation}
  \label{forcesnear}
  \begin{split}
    F(x,y,z)
    &=\,\tilde F(x,y,z) +
    \begin{cases}
      \frac{1}{(x^2+y^2+z^2)^{\frac{3}{2}}}
      \left(x,y,z\right)^T & (x,y,z)\neq(0,0,0)\\
      0 & (x,y,z)=(0,0,0)
    \end{cases}
  \end{split}
\end{equation}
where
\small
\begin{equation*}
  \begin{split}
    \tilde F_x(x,y,z)=\,
    &  8\pi u_x^2\sum_{l,p>0}p\left(\KN(\omega_p\rho_l) +
      \KN(\omega_p\rho_{-l})\right)\sin(\omega_p x) +\\
    &  \sum_{k=1}^{N_\psi-1}\left(\frac{x+k\lx}{r_{k}^3} +
      \frac{x-k\lx}{r_{-k}^3}\right) +\\
    &  u_x^2\sum_{n\ge 0}\binom{-\frac{1}{2}}{n}\frac{\left(
        \psi^{(2n+1)}(N_\psi + u_x x) -
        \psi^{(2n+1)}(N_\psi - u_x x)\right)}{(2n)!}(u_x\rho)^{2n}\mcm\\
  \end{split}
\end{equation*}
\begin{equation*}
  \begin{split}
    \tilde F_y(x,y,z)=\,
    &  8\pi u_x^2\sum_{l,p>0} p \biggl(
    \frac{(y+l\ly)\KO(\omega_p\rho_l)}{\rho_l}\,+
    \frac{(y-l\ly)\KO(\omega_p\rho_{-l})}{\rho_{-l}}
    \biggl)\cos(\omega_p x) -\\
    &  4\pi u_y u_x\sum_{n\ge 1}\frac{b_{2n}}
    {(2n)!}\Im\bigl((2\pi u_y (z+iy))^{2n-1}\bigr) +
    \sum_{k=1}^{N_\psi-1}\left(\frac{y}{r_{k}^3} +
      \frac{y}{r_{-k}^3}\right) +\\
    &  u_x^3 y\sum_{n\ge 1}\binom{-\frac{1}{2}}{n}\frac{\left(
        \psi^{(2n)}(N_\psi + u_x x) +
        \psi^{(2n)}(N_\psi - u_x x)\right)}{(2n-1)!}(u_x\rho)^{2(n-1)}\mcm
  \end{split}
\end{equation*}
\begin{equation*}
  \begin{split}
    \tilde F_z(x,y,z)=\,
    &  8\pi u_x^2\sum_{l,p>0} p \left(
      \frac{z\KO(\omega_p\rho_l)}{\rho_l} +
      \frac{z\KO(\omega_p\rho_{-l})}{\rho_{-l}}
    \right)\cos(\omega_p x) -\\
    &  4\pi u_y u_x\sum_{n\ge 1}\frac{b_{2n}}{(2n)!}
    \Re\bigl((2\pi u_y (z+iy))^{2n-1}\bigr) +
    \sum_{k=1}^{N_\psi-1}\left(\frac{z}{r_{k}^3} +
      \frac{z}{r_{-k}^3}\right) +\\
    &  u_x^3 z\sum_{n\ge 1}\binom{-\frac{1}{2}}{n}\frac{\left(
        \psi^{(2n)}(N_\psi + u_x x) +
        \psi^{(2n)}(N_\psi - u_x x)\right)}{(2n-1)!}(u_x\rho)^{2(n-1)}\mcm
  \end{split}
\end{equation*}
\normalsize
where $K_1$ is the modified Bessel function of order one.
%%%%%%%%%%%%%%%%%%%%%%%%%%%%%%%%%%%%%%%%%%
\section{Error Estimates}\label{sec:errors}
%%%%%%%%%%%%%%%%%%%%%%%%%%%%%%%%%%%%%%%%%%
For an implementation which is of practical use, error estimates are needed.
Since we calculate the energy (forces) by summing up pairwise energies $E_i$
(their differential), it is reasonable to derive an upper bound for the error
of $E$. The maximal pairwise error $\epsilon_E$ for the energy is induced by
the calculation of the formulas \eqref{fullfar} and \eqref{fullnear} with
finite cutoffs. Likewise we can derive an maximal pairwise error $\epsilon_F$
for the forces by formulas \eqref{forcesfar} and \eqref{forcesnear}, and one
can furthermore give an upper bound for the commonly used RMS--error of the
forces.  We will show later that the error distribution for \algo{MMM2D} is
highly non--uniform, and leads typically to an RMS--error is much lower than
our bound. Thus the RMS--error is not the optimal error measure.
Furthermore our implementation shows that an increase
in precision has little impact on the calculation time of \algo{MMM2D}, quite
contrary to mesh based\cite{deserno98a} and other methods\cite{widmann97a}.
\subsection{Error of the far formula}
For the far formula given by \eqref{fullfar} and \eqref{forcesfar},
respectively, 
we use a radial cutoff. So the summation is not performed
over all $(p,q)\neq 0$, but only for those $(p,q)\in\Gamma_R$
where
\begin{equation}
  \label{defgammar}
  \Gamma_R=\left\{(p,q)\in [0,\infty)^2 \backslash \{0\} \,|\,
    u_x^2(p-1)^2+u_y^2(q-1)^2 < R^2\right\}\mpt
\end{equation}

By a simple approximation of the sums by integrals it is easy to
derive upper bounds. For the potential $\phi$ we find
the estimate
\begin{equation}
  \label{errorfarp}
  \tau_E:=\left(1+\frac{u_x+u_y}{\pi R}\right)\frac{e^{-2\pi R|z|}}{|z|}\mpt
\end{equation}
This upper bound is \emph{not} valid for the forces. For all three
components $F_x$, $F_y$ and $F_z$ of the forces \emph{and} the
potential we find the upper bound
\begin{equation}
  \label{errorfarpf}
  \tau_F=\frac{e^{-2\pi R|z|}}{|z|}\left(2\pi R +
    2 (u_x + u_y) + \frac{1}{|z|}\right)\mpt
\end{equation}
This value is precise only for $F_z$. The other force components
and the potential show even a better convergence. Common to both error
formulas is the {\it exponential decrease} of the error with $R|z|$.

\subsection{Error of the near formula}
The near formula given by equations \eqref{fullnear} rsp.
\eqref{forcesnear} contains three sums with different cutoffs.

For the first sum containing Bessel functions it is reasonable
to sum over all $(p,q)\in\Omega_L$ where
\begin{equation}
  \begin{split}
    \Omega_L:=\left\{(p,l)\,\bigg|\quad
      0 < p < \frac{L}{\pi u_x}\quad\text{und}\quad
      0 < l < \frac{L}{\omega_p} + 1\quad\right\}\mpt
  \end{split}
\end{equation}
If $L\ge \max(3u_y,\pi u_x + u_y)$, $|x|\le\lx/2$, $|y|,|z|\le\ly/2$,
the inequality $K_0(x)\le e^{-x}$ for $x\ge 3$ can be used to derive
the upper bound
\begin{equation}
  \begin{split}
    \tau_B=8u_x\max(2\pi u_x, 1)e^{-\ly L}\Biggl(&
    \frac{e^{\pi u_x \ly}}{\pi u_x \ly}
    \left(\frac{L + u_y}{\pi u_x} - 1\right)+
    \sum_{p=1}^{\lceil\frac{L}{\pi u_x}\rceil - 1}
    p e^{-\pi u_x \ly p}
    \Biggr)
  \end{split}
\end{equation}
again by an approximation of the sum by an integral.
The Gaussian bracket $\lceil\cdot\rceil$ denotes rounding up.
This is an estimate for the absolute error of all force
components and the potential.

For the second sum in \eqref{fullnear} containing the Bernoulli
numbers we use the estimate $|b_{2n}|\le 4\frac{(2n)!}{(2\pi)^{2n}}$
from \cite{abramowitz} to obtain the overall error estimate
\begin{equation}
  \label{errbernsum}
  \tau_K= 16 u_x u_y (u_y\rho)^{2N-1}\le 16\sqrt{2} u_x u_y 2^{-N}\mpt
\end{equation}
for the summation of this sum up to $N$. Note that this sum does not
contribute to $F_x$, due to the artificially broken symmetry in the
derivation. The error estimate contains the particle position dependent term
$\rho$. By using a table lookup scheme for $N(\rho)$ one can chose the
appropriate cutoff at runtime to speed up the calculation.  

For the last sum containing the polygammma functions there is no error
estimate necessary. The terms in this sum drop monotonously with
alternating signs and therefore the absolute value of the last added
term is an upper bound on the error. So the cutoff is determined at
runtime.

\section{Efficient Implementation}\label{sec:implementation}
In this section we want to describe how the formulas \eqref{fullfar} and
\eqref{fullnear} can be efficiently implemented to achieve the time scaling of
$\O(N^{\frac{5}{3}}\log(N)^2)$.

The simulation box is split into $B$ equally sized slices along the
$z$--axis. We will see later how the parameter $B$ should be chosen.
Slice $S$ contains all particles $i$ that fulfill
$\frac{S}{B} < \frac{z_i - z_{min}}{\lz}<\frac{S+1}{B}$, where
$z_{min}$ and $z_{max}$ are respectively the minimal and maximal
$z$--coordinates. Furthermore we set $\lz= z_{max}-z_{min}$ and
$I_S$ should be an arbitrary enumeration of the particles in
slice $S$.

Now for all particles in slice $S$ the interaction (potential or force)
with the particles in the slices $S-1$, $S$  and $S+1$ (if existent)
will be calculated using the near formula \eqref{fullnear} rsp.
\eqref{forcesnear}. For the other slices the far formula
\eqref{fullfar} rsp. \eqref{forcesfar} is used. In the following
picture the splitting of the calculation for all particles in the
black slice is shown:
\begin{center}
  \includegraphics[height=3cm]{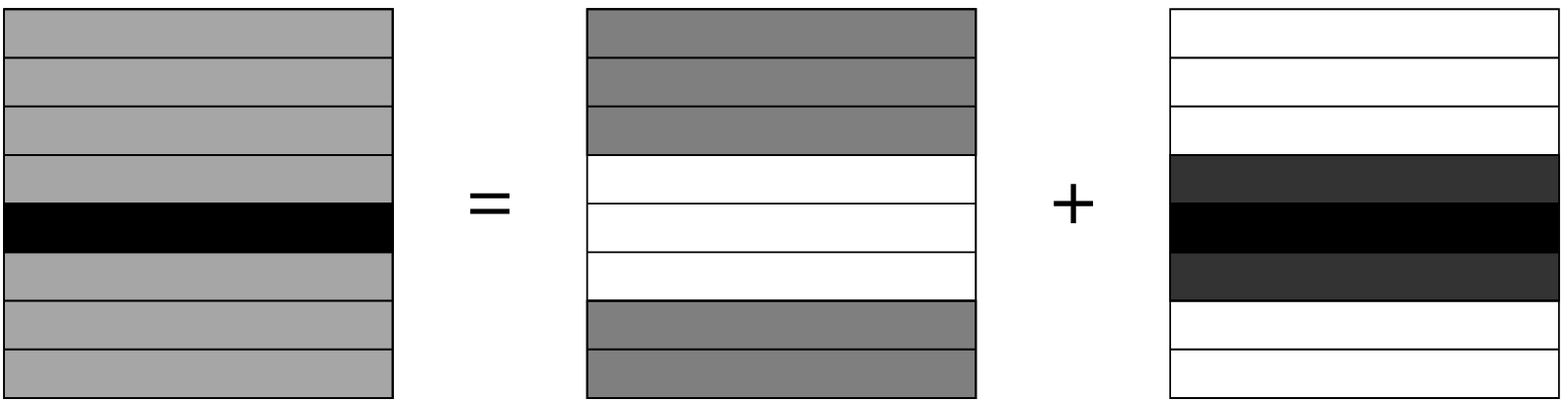}
\end{center}

For the the near formula to be valid we need $|z_i-z_j|\le\ly/2$ for
particles $i$ and $j$ located in adjacent slices. This gives the
constraint
\begin{equation}
  \label{minB}
  \frac{2\lz}{B}\le\frac{\ly}{2}
\end{equation}
Therefore $\ly$ should be as large as possible and it might be useful to
exchange the $x$-- and $y$--axis if $\lx>>\ly$. In typical cases this will not
be necessary since $B$ is normally chosen large enough to yield a minimal computation
time.

The minimal distance of two particles that are treated by the far
formula is $b=\lz/B$. With this distance it is easy to use the error
estimate \eqref{errorfarp} rsp. \eqref{errorfarpf} for the far formula
to calculate the cutoff $R$. Essentially we have
\begin{equation}
  \label{R0dep}
  R\sim\frac{B}{\lz}\log\left(\frac{B}{\epsilon\;\lz}\right)\mpt
\end{equation}

\subsection{Partial ordering of the particles}
First the sets $I_S$ are determined. This can be done in time
$\O(N)$, as the following algorithm shows:
{\samepage
  \begin{prog}[Partial sorting]
    \label{progsort}
    \hspace{\fill}
    \begin{list}{}{}
    \item For all particles $j$ determine its slice $S_j$ as
      \begin{equation*}
        S_j=\left\lfloor
          B\frac{z_i - z_{min}}{z_{max}-z_{min}}\right\rfloor\mpt  
      \end{equation*}
      Add $j$ to $I_{S_j}$.
    \end{list}
  \end{prog}}
The Gaussian bracket $\lfloor\cdot\rfloor$ denotes rounding down.
It is more efficient to reorder the particle list so that
the particles of a given slice are stored adjacently, since later
many indirect accesses via the sets $I_S$ can be omitted.

\subsection{Implementation of the far formula}
The calculation of the far formula consists of summing terms with
frequencies $(p,q)$ in the Fourier space.
Now we show how the calculation of the contribution to the
energies $E_i$ by a fixed frequency $(p,q)$ can be done such
that the computing time is $\O(N)$.
The algorithm presented can easily be adapted for the calculation
of the sums with only a single Fourier frequency and the $|z|$--sum.
The calculation of the forces can be done simultaneously with the
calculation of the potential using the same expressions.
 
In the beginning we concentrate on a single particle $i$ located in slice
$S_i$. The far formula is used to calculate the contributions from all
particles in non--adjacent slices, i.\ e. slices $S_j$ where
$|S_j-S_i|>1$. First we restrict attention to the particles $j$
in slices $S_j < S_i-1$, that is, to the particles in the set
$\bigcup_{S<S_i-1}I_S$. The cosine terms are separable via the
addition theorem and since $|z_i-z_j|=z_i-z_j$ for these slices, we have 
\begin{equation}\label{eq:sins}
  \begin{split}
    &\sum_{\substack{j\in I_S\\S < S_i - 1}}
    q_iq_j\frac{e^{-2\pi f_{pq}|z_i-z_j|}}{f_{pq}}
    \cos(\omega_p (x_i - x_j))\cos(\omega_q (y_i - y_j)) =\\
    &q_i\frac{e^{-2\pi f_{pq}z_i}}{f_{pq}}
    \cos(\omega_p x_i)\cos(\omega_q y_i)
    \sum_{\substack{j\in I_S\\S < S_i - 1}}q_je^{2\pi f_{pq}z_j}
    \cos(\omega_p x_j)\cos(\omega_q y_j) + \\
    &q_i\frac{e^{-2\pi f_{pq}z_i}}{f_{pq}}
    \cos(\omega_p x_i)\sin(\omega_q y_i)
    \sum_{\substack{j\in I_S\\S < S_i - 1}}q_je^{2\pi f_{pq}z_j}
    \cos(\omega_p x_j)\sin(\omega_q y_j)
    + \\
    &q_i\frac{e^{-2\pi f_{pq}z_i}}{f_{pq}}
    \sin(\omega_p x_i)\cos(\omega_q y_i)
    \sum_{\substack{j\in I_S\\S < S_i - 1}}q_je^{2\pi f_{pq}z_j}
    \sin(\omega_p x_j)\cos(\omega_q y_j)
    + \\
    &q_i\frac{e^{-2\pi f_{pq}z_i}}{f_{pq}}
    \sin(\omega_p x_i)\sin(\omega_q y_i)
    \sum_{\substack{j\in I_S\\S < S_i - 1}}q_je^{2\pi f_{pq}z_j}
    \sin(\omega_p x_j)\sin(\omega_q y_j)\mpt
  \end{split}
\end{equation}

The sum for the $S > S_i + 1$ is very similar, just the sign of the
$z_i$ and $z_j$ is exchanged. For all particles $j$ only the eight
terms
\begin{equation*}
  \chi^{(\pm,s/c,s/c)}_j= q_je^{\pm 2\pi f_{pq}z_j}
  \sin/\cos(\omega_p x_j)\sin/\cos(\omega_q y_j)
\end{equation*}
are needed. The upper index describes the sign of the exponential term
and whether sine or cosine is used for $x_j$ and $y_j$ in the obvious way.
These terms can be used for all expressions on
the right hand side of Eq. \eqref{eq:sins}. Moreover it is easy to see from
the addition theorem for the sine function that these terms also
can be used to calculate the force information up to simple prefactors
that depend only on $p$ and $q$. Once these terms are calculated
for every particle,
the rest of the calculation only consists of adding products of these
terms $\chi^{(\pm,s/c,s/c)}_j$ with correct prefactors to the potential
and the force. While it is easy to find these prefactors from the above
formulas, it requires some care to insert the correct signs from
the addition theorems into an implementation. Again we
present the algorithm in pseudo code:

\begin{prog}[Calculation of the far formula for a fixed $(p,q)$]
  \label{progfar}
  \hspace{\fill}
  \begin{enumerate}
  \item For all particles $j$ calculate
    $\chi^{(\pm,s/c,s/c)}_j$.
  \item For all slices $l$ calculate
    \begin{equation*}
      \tilde\chi^{(\pm,s/c,s/c)}_l=
      \sum_{j\in I_l}\chi^{(\pm,s/c,s/c)}_j\mpt
    \end{equation*}
  \item For all slices $k$ calculate
    \begin{align*}
      \xi^{(+,s/c,s/c)}_k=\,
      \sum_{\substack{j\in I_S\\S<k-1}}\chi^{(+,s/c,s/c)}_j
      =\,\sum_{l<k-1}\tilde\chi^{(+,s/c,s/c)}_l \\
      \xi^{(-,s/c,s/c)}_k=\,
      \sum_{\substack{j\in I_S\\S>k+1}}\chi^{(-,s/c,s/c)}_j
      =\,\sum_{l>k+1}\tilde\chi^{(-,s/c,s/c)}_l\mpt
    \end{align*}
  \item For all particles $i$ add the appropriate linear combination
    of the terms $\chi^{(\mp,s/c,s/c)}_i$ (left hand terms)
    and the terms $\xi^{(\pm,s/c,s/c)}_{S_i}$ (right hand terms)
    to $E_i$ and $F_i$. Remember that a term with upper index
    $+$ is always combined with a term of upper index $-$ and vice versa.
  \end{enumerate}
\end{prog}

There are some tricks to speed up this algorithm. An easy way is to
pre-calculate the sine and cosine terms for several frequencies $p$ and
$q$. And now it becomes clear that these loops are more efficient, as
mentioned above, when particle data and results are stored
consecutively in their respective arrays.

If larger cutoffs $R$ are needed, the terms $e^{f_{pq}z_i}$ may get
large rapidly. Then it might be useful to use the scaled coordinates
$e^{f_{pq}(z_i-z_b)}$ instead, where $z_b$ is constant for a slice
(e.g. the upper, lower border or the center). Then the terms
$\chi$ for the slices are calculated as before, the scaling is
corrected during the summation for the $\xi$. The advantage of this is
that the expression $e^{f_{pq}(z_b-z_{\tilde{b}})}$ is numerically much
more stable than $e^{f_{pq}z_b}e^{-f_{pq}z_{\tilde{b}}}$.

Remember that this algorithm has to be executed for every frequency
$(p,q)$ and also for the single frequency sums and for the $|z|$--sum.

\subsection{Calculation of the near formula}
The calculation is algorithmically easy: It consists of two nested
loops:
\begin{prog}[Calculation of the near formula]
  \label{prognear}
  \hspace{\fill}
  \begin{list}{}{}
  \item  For every slice $S$ let $I_S=\{i_1,\hdots,i_l\}$. Then
    for $i_k=i_1,\hdots,i_l$ add the self energy, i.\ e. the
    potential, that it feels from its own copies
    $q_{i_k}^2\tilde\phi(0,0,0)$.
    
    Furthermore calculate the potential and the force the particle $i_k$
    feels from the particles $j=i_{k+1},\hdots,i_l$ and the particles
    $j\in I_{S+1}$. Add these terms to the potential and the force of
    \emph{both} particles, changing the sign of the force for
    particle $j$.
  \end{list}
\end{prog}

The self-energy can be calculated by formula \eqref{fullnear}.
For $\lx=\ly=1$ it is $\tilde\phi(0,0,0)\approx-3.90026$.

Now we want to give some hints for the calculation of the somewhat
unusual functions involved. For the Bessel functions one can find
approximations for example in \cite{abramowitz}. Code using high
order Chebychev approximations to calculate the Bessel functions
is freely available, for example from the GNU scientific
library~\cite{gsl}.

Also we need a table of the Bernoulli numbers, which is
easy to obtain. But it is more efficient to store the terms
\begin{equation*}
  \frac{b_{2n}}{2n (2n)!}(2\pi)^{2n}\mpt
\end{equation*}
The calculation of the sum over the Bernoulli numbers
can be accelerated by making the cutoff depend on
$u_y\rho$ using the error formula \eqref{errbernsum}.

The polygamma functions needed for the last sum are not calculated directly,
instead we will give quickly converging power series for the functions
\begin{align}
  \label{modpsi}
  \tilde\psi_N^{(2n)}(x)&:=\frac{1}{(2n)!}
  \left(\psi^{(2n)}(N+x)+\psi^{(2n)}(N-x)\right)
  \quad\text{and}\\
  \tilde\psi_N^{(2n+1)}(x)&:=\frac{1}{2n!}
  \left(\psi^{(2n+1)}(N+x)-\psi^{(2n+1)}(N-x)\right)\mcm
\end{align}
since these functions have a much better convergent series expansion
than the polygamma functions.

Letting
\begin{equation*}
  \tilde\psi_c^{(2n)}(x)=\binom{-1/2}{n}\tilde\psi_{N_\psi}^{(2n)}(x)
  \quad\text{and}\quad
  \tilde\psi_c^{(2n+1)}(x)=\binom{-1/2}{n}\tilde\psi_{N_\psi}^{(2n+1)}(x)
  \mcm
\end{equation*}
the terms needed in the last sum can be calculated as
\begin{align*}
  u_x\tilde\psi_c^{(2n)}(x)(u_x\rho)^{2n}         \mcm\quad
  &u_x^2\tilde\psi_c^{(2n+1)}(x)(u_x\rho)^{2n}     \mcm\\
  2nyu_x^3\tilde\psi_c^{(2n)}(x)(u_x\rho)^{2(n-1)}\mcm\quad
  &2nzu_x^3\tilde\psi_c^{(2n)}(x)(u_x\rho)^{2(n-1)}\mpt
\end{align*}

For the calculation of the power series we need the
generalized zeta function or \emph{Hurwitz zeta function}
\begin{equation*}
  \zeta(m, N):=\sum_{l=N}^\infty l^{-m}\mpt
\end{equation*}
Algorithms to calculate this function can be found in
\cite{abramowitz}, but code calculating this function is also freely
available, again from the GNU scientific library~\cite{gsl}.

Again from \cite{abramowitz} we obtain with a little effort the
series expansions
\begin{equation}
  \begin{split}
    \tilde\psi_N^{(2n)}(x) &= -2\sum_{k=0}^\infty\binom{2n+2k}{2k}
    \zeta(2n+2k+1, N)x^{2k}\mcm\\
    \tilde\psi_N^{(0)}(x) &= -\gamma -
    2\sum_{k=1}^\infty\zeta(2k+1,N)x^{2k}\mcm\\
    \tilde\psi_N^{(2n+1)}(x) &= 2 x
    \sum_{k=0}^\infty\binom{2n+2k+1}{2k+1}
    \zeta(2n+2k+2, N)x^{2k}\mpt
  \end{split}
\end{equation}

\subsection{Computation time}
Using the implementation scheme given above, an estimation of the
asymptotic computation time can be derived. We assume that the
particles are distributed uniformly in a box. In other words, we
assume that $N/B$ particles are located in every slice.

The program~\ref{progsort} has clearly a computation time of $\O(N)$.

The program~\ref{progfar} is executed for all frequencies
$(p,q)\in\Gamma_{R}$, that is $\O({R}^2)$ times.
The first two steps of the program obviously have a linear computation
time, the third step a computation time of order $\O(B^2)$. The last
step is linear again.
In total we obtain a time needed for the program of
$\O(R^2N)+\O(R^2B^2)$.

The time for the calculation of the near formula for a
single particle pair is assumed to be constant. This is sensible since
at least the time for the sum over the Bessel functions is constant
and the times for the other two sums are bounded. Therefore
the time needed for program~\ref{prognear} is $\O(N(2N/B))$.

If we insert $R\sim B/\lz\log(B/(\epsilon\;\lz))$ from equation
\eqref{R0dep}, we obtain an asymptotical computation time of
\begin{equation}
\label{asymaufw}
\O(B^2\log(B)^2N)+\O(B^4\log(B)^2)+\O(N^2/B)+\O(N)\mpt
\end{equation}
for the complete program.

Choosing $B\sim N^{\frac{1}{3}}$, we obtain a computation time of
\begin{equation*}
  \O(N^{\frac{5}{3}}\log(N)^2)+\O(N^{\frac{4}{3}}\log(N)^2)+
  \O(N^{\frac{5}{3}})\mpt  
\end{equation*}
This gives a total computational time of
$\O(N^{\frac{5}{3}}\log(N)^2)$, which is easily seen to be optimal
from equation \eqref{asymaufw}.

Although we know the optimal proportionality of $B$, this parameter
has to be tuned to the underlying hardware. But this can be done by a
simple trial and error algorithm at runtime, since the computational
error does not depend on $B$.

%%% Local Variables: 
%%% mode: latex
%%% TeX-master: "article1"
%%% End: 

\section{Numerical Examples}\label{sec:numerics}
In this section we give some numerical results from our implementation
of the \algo{MMM2D} algorithm. The computations were performed
on a Compaq XP1000 with a 667 MHz EV67--CPU.

\subsection{Regular Systems}
First we want to demonstrate that \algo{MMM2D}
reproduces some known results from Ref.\cite{widmann97a}. We also give the
consumed computation times for \algo{MMM2D} and for an implementation
of the $2d$--Ewald method.

We consider the following systems, which are in more detail described
in Ref.\cite{widmann97a}:
\begin{trivlist}
\item[System 1:] 100 particles distributed regularly in a square of length
  $L=1$. The particles have a unit charge value
  with chessboard--like alternating signs.
\item[System 2:] The same as above, but one particle elevated by $0.5$
  out of the plane.
\item[System 3:] This system consists of 25 particles in a square and
  one elevated by $0.2$ out of the plane in the center of the square.
  The charges in the plane are again chessboard--like distributed, the
  full system is charge neutral.
\end{trivlist}

Table~\ref{tabwid} shows some results for these systems. For the
numerical values, we tuned \algo{MMM2D} for a maximal pairwise error
of $\epsilon=10^{-6}$ in the forces by the error formulas given above.
For the timings $T_{\text{\algo{MMM2D}}}$ we used a maximal pairwise
error of $\epsilon=10^{-4}$ (timings for $\epsilon=10^{-6}$ in
parentheses).  The $2d$--Ewald method was tuned for a RMS--error of
$10^{-4}$, since for a typical simulation an error of $10^{-4}$ is
sufficient. Note that for \algo{MMM2D} the {\it actual} RMS--errors
were $10^{-7}$ for $\epsilon= 10^{-4}$ and $10^{-9}$ for $\epsilon=
10^{-6}$. For system 1 the forces on all particles are equal in
magnitude and the force on an arbitrary particle is given, for Systems
2 and 3 the forces on the elevated particle are given.

\begin{table}[htbp]
  \begin{center}
\caption{Force and energy results and computation times
      for the systems 1, 2 and 3}
    \fontsize{9}{11}
    \selectfont
    \begin{tabular}[c]{l|c|c|c|c|c|c}
      System & $F_x$ & $F_y$ & $F_z$ & $E_{total}$ &
      $T_{\text{\algo{MMM2D}}}$ &
      $T_{2d-\text{Ewald}}$ \\
      & & & & & [ms] & [ms] \\
      \hline
      1 & $5\cdot 10^{-9}$   & $2\cdot 10^{-8}$   &   0
      & -807.771313 &            74.1(110)  & 318   \\
      % 1e-6 timings
      % for XP 1000             110    & 317
      % for AMD   TB   1333      73.5  & 219
      % for Intel PIII  900     140    & 575
      2 & $-2\cdot 10^{-17}$ & $-2\cdot 10^{-17}$ &  -7.765381
      & -792.588065 &            72(108)    & 317   \\
      % for XP 1000             108    & 317
      % for AMD   TB   1333      72.5  & 219
      % for Intel PIII  900     139    & 579
      3 & $4\cdot 10^{-16}$  & $1\cdot 10^{-15}$  & -10.364160
      &  -86.565859 &             4.91(7.32) &  21.1 \\
      % for XP 1000               7.32 &  20.9
      % for AMD   TB   1333       4.8  &  14.7
      % for Intel PIII  900       8.85 &  38.5   

    \end{tabular}
    
    \label{tabwid}
    \normalsize
  \end{center}
\end{table}

In system 1 the $z$--distance is 0 for every pair of particles, and therefore
only the near formula is used. For the systems 2 and 3 only the far formula
was used for the interactions with the elevated particle, and the
precision for those systems is much better than for the first, as can be
inspected in Tab.~\ref{tabwid}.  This is due to the exponential factor
containing the $z$--distance in the error formulas Eqs. \eqref{errorfarpf} and
\eqref{errorfarp}. This factor leads to a non--uniform error
distribution with respect to $z$ and forces \algo{MMM2D} to be
extremely precise for some particle positions, and therefore the
forces of the elevated particles are calculated with likewise
precision. We will show this in more detail in the next part.
The computation times will be discussed later.

\subsection{2--particle systems}\label{subsec:2particle}

\begin{figure}[bthp]
  \begin{center}
    \includegraphics[width=\picturewidth]{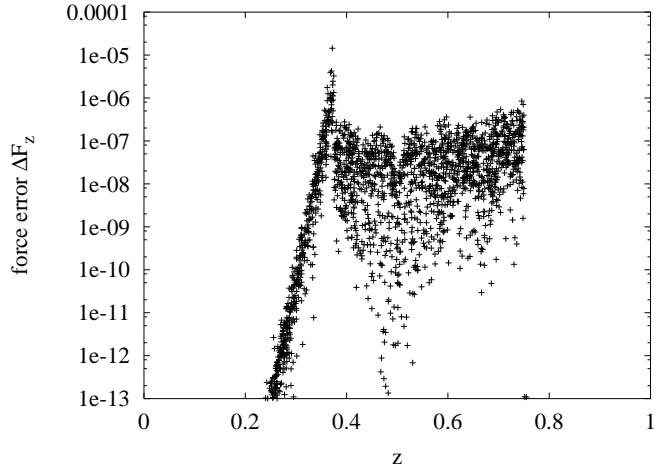}
    \caption{Absolute $z$--force error in dependence of the
      $z$--distance of the two-particle system (compare text)}
    \label{picerr}
  \end{center}
\end{figure}

For Figure~\ref{picerr} we used two particles, one located at $(0,0,0.5)$, the
other randomly placed in a cubic box of unit length. We show the error in the
$z$--force on the fixed particle for a calculation with $B=8$ and a maximal
pairwise error of $\epsilon=10^{-6}$.

Since the box has a height of 1 and $B=8$, the slices have a height of
$0.125$ and the forces are calculated with the near formula if and
only if $0.375<=z<0.750$. In this distance range is the maximum error
nearly constant and scatters well below the maximal pairwise error.
The additional structure which is observable in the error distribution is due
to the way the different sums in the near formula \eqref{fullnear} are
truncated depending on $z$ in order to reach the prescribed precision.

For $z<0.375$ the exponential drop of the error predicted by the far error
formulas Eqs.\eqref{errorfarpf} and \eqref{errorfarp} is clearly visible. This
effect cannot be seen on the right side for large $z$ since the error of the
far formula is symmetrical, while the distance where the near formula is used
is asymmetrical with respect to the fixed particle. If the fixed particle
would be located at the upper bound of its slice, for example at
$(0,0,0.624)$, the far error would be visible on the right side, but not on
the left.

The exponential drop of the far formula error suggests that one could reduce
the numbers of added Fourier components for more distant slices to reduce the
computation time. However, this would only save a number of operations (additions and
multiplications) of the order of the number of slices, and the overhead for
singling out the slices compensates the gain in time.

\begin{figure}[bthp]
  \begin{center}
    \includegraphics[width=\picturewidth]{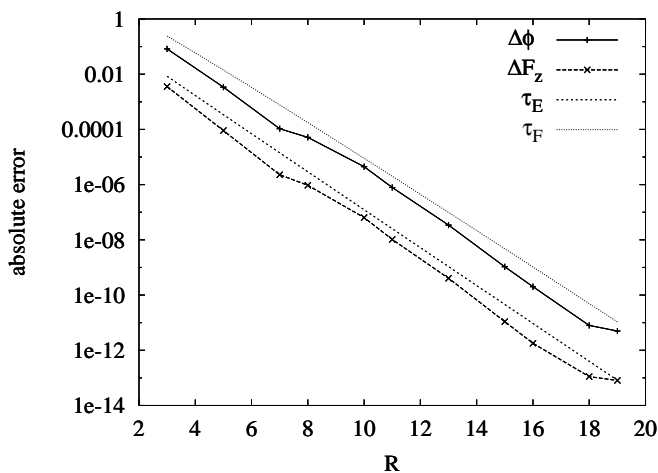}
    \caption{Maximal absolute errors of the $z$--force and the energy for the
      2--particle system (compare text)}
    \label{picerrform}
  \end{center}
\end{figure}

In the time scale considerations above, the error estimate for the far
formula has a central place. Therefore we want to show that the
estimates $\tau_F$ (for the forces) and $\tau_E$ (for the energy) are
accurate. Figure~\ref{picerrform} shows the error estimates and the
maximal errors in dependence of the cutoff $R$. The system
investigated here again consists of two particles, one fixed at
$(0,0,0)$, the other one randomly placed in
$[0,1]\times[0,1]\times[0.25,1]$. It can be seen that both error
formulas overestimate slightly the error, but are sufficiently close
to the numerical results.

\subsection{Randomly distributed systems}
Now we want to investigate the RMS--error scaling of \algo{MMM2D} and the
computation time for larger systems. The systems investigated consist of $N$
particles placed randomly in a cubic simulation box of unit length. Half of
the particles carry a charge of $1$, half of them a charge of $-1$. For the
calculations \algo{MMM2D} was tuned to a maximal pairwise error of
$\epsilon=10^{-4}$.

\begin{figure}[bthp]
  \begin{center}
    \includegraphics[width=\picturewidth]{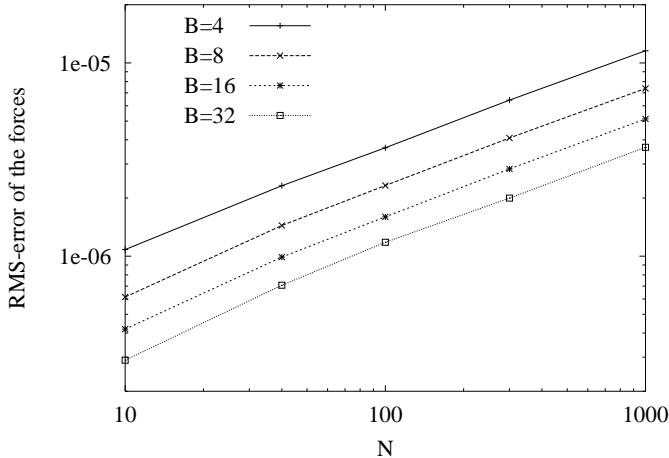}
    \caption{RMS--error of the forces calculated by \algo{MMM2D}
      for different values of $B$ in dependence of the particle
      numbers $N$ for fixed $\epsilon = 10^{-4}$}
    \label{picerrN}
  \end{center}
\end{figure}

The error distribution is highly non--uniform as we have shown before.
Therefore it is more safe to use the error formulas given above to constrain
the maximal pairwise error rather than the RMS--error, as is done usually.
Nevertheless we investigated the actual RMS--error for the uniformly random
systems described above.  Figure~\ref{picerrN} shows the RMS--error for
different particle numbers and slice numbers. By the arguments given in Sec.
\ref{subsec:2particle}, the error drops when more slices are used.

If one assumes a standard deviation of the pairwise error of $\sigma$,
and furthermore that the pairwise errors are independent, which is true for
random systems, it is easy to see that the RMS--error has an expectation value
of $\frac{Q^2}{\sqrt{N}}\sigma$ \cite{deserno98b}, with $Q^2 = \sum_i
q_i^2$, which can be seen in Figure~\ref{picerrN}. Thus our RMS-error behaves
like the ones for the usual Ewald methods \cite{deserno98b}. Furthermore we
have $\sigma \le \epsilon$ so that the RMS--error can be bounded using our
error estimates. Note that again the RMS-error is much smaller than the
maximal pairwise error (for 100 particles, $3\cdot 10^{-6}$ instead of
$10^{-4}$), although the gain is smaller than for the highly structured
systems 1-3.

\begin{figure}[bthp]
  \begin{center}
    \includegraphics[width=\picturewidth]{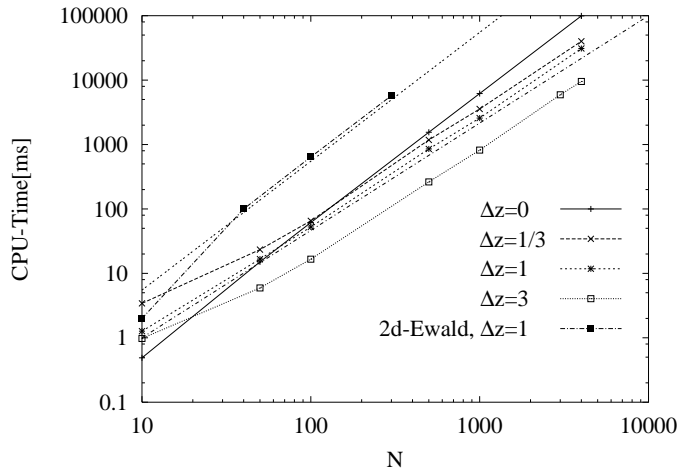}
    \caption{Optimal computation times for different particle
      numbers $N$.}
    \label{pictN}
  \end{center}
\end{figure}

Having laid the theoretical foundations of the time scaling in
Sec.~\ref{sec:implementation}, we want to show that the theoretical
scaling can be achieved in a real computation. Figure~\ref{pictN}
shows the optimal computation time depending on the particle number
$N$ for the systems considered above. We also give data for
different box heights $\Delta z$. The $x$ and $y$ sides are still
fixed to unit length. The pairwise error in the forces was fixed to
$\epsilon=10^{-4}$, resulting in the better RMS-errors given above.
The optimal number of slices $B$ depends on $\Delta z$. For example,
for $\Delta z=0$ we always have $B=1$ and a purely $\O(N^2)$
algorithm, as is seen in the figure. For $\Delta z=1$ the optimal
slice numbers were $B=4$ for $N\le 100$, $B=8$ for $N\le 1000$ and
$B=16$ for $N=4000$, and the $\O(N^{5/3})$ scaling is achieved. It can
also be seen that \algo{MMM2D} becomes faster with growing $\Delta z$,
as one expects from the error formula.
For comparison also computation times for the $2d$--Ewald method (at
$\Delta z=1$) and the theoretical scaling curves of $\O(N^{5/3})$ and
$\O(N^2)$ are included.

The computation times show that \algo{MMM2D} is always faster than the
HBC method, even for small particle numbers (although we achieve
a much smaller RMS--error). This can also be
seen in the computation times given in Tab.\ref{tabwid}. These
computation times also demonstrate that increasing the precision does not
massively impact the computation time of \algo{MMM2D}. This is due to
the exponential error decrease with respect to the cutoff $R$.
Since System 1 in this table is purely two dimensional, \algo{MMM2D}
only uses the near formula for the calculation. Therefore even our
near formula is superior to the $2d$--Ewald method.

Since \algo{MMM2D} and the algorithm described by Smith \cite{smith88a} are
similar up to the difference that Smith uses the $2d$--Ewald method instead of
our near formula, \algo{MMM2D} is also faster than the algorithm of Smith.
From the numbers given by Widmann \cite{widmann97a} one can furthermore
conclude that \algo{MMM2D} is also faster than the algorithm proposed by
Hautman and Klein\cite{hautman92a}, although we did not test this method
explicitly.

\algo{MMM2D} is also similar in speed at our accuracy with respect to the two
recent Ewald-type algorithms proposed in Refs.  \cite{kawata01a,kawata01b}.
However, these two methods also have the drawback that no error estimates
exist, and these will also be hard to develop because of the numerical
integration involved.  Furthermore it seems unreasonable that their algorithm
gets slower for larger $\Delta z$, while \algo{MMM2D} gets faster, and that
their algorithm is fastest for the largest tested real-space
cutoffs\cite{kawata01a} although they claim that their new Fourier-space
expression is responsible for the increase in speed. On the basis of their
numbers and non-comparable test cases we are unable to draw further
conclusions.

\section{Conclusion}\label{sec:conclusion}
We developed a new method termed \algo{MMM2D} to accurately calculate the
electrostatic energy and forces on charges being distributed in a two
dimensional periodic array of finite thickness ($\ddh$ geometry). It has the
advantage of requiring only a computational time of $\O(N^{5/3})$ instead of a
quadratic scaling as previous methods show. It is based on a convergence factor
approach and as such does not require any fine--tuning of any Ewald parameters
for convergence. Moreover, we derived rigorous error bounds and numerically
verified them on specific examples.

Our method is not only asymptotically faster than all $2d$--Ewald
methods, but is about four times faster than the HBC method for small
numbers of particles, and should also be superior to the method of
Hautman and Klein, and to the new methods of Refs.
\cite{kawata01a,kawata01b} at our accuracies.  Our energy and force
results achieve a maximal pairwise error of $10^{-4}$ with minimal cutoffs,
and $10^{-8}$ can easily be achieved without large increase in
computational time.  Therefore our method is even more superior
whenever a high accuracy is needed.  Although the presented formulas
may look awkward at the beginning, once implemented, the method is
easy to use because the cutoffs for the sums can be easily determined
at runtime. Only the number of slices $B$ must be optimized for speed,
but, although not implemented, this parameter could also be determined
during execution.  This is in contrast to basically all $2d$--Ewald
methods, for which the tuning of the various parameters is very
crucial and error estimates do not exist.

The numerical code for \algo{MMM2D} in C++ can be requested from the
authors. This code also provides interfaces for use with FORTRAN and C.

\section*{Acknowledgements}
We would like to thank J. DeJoannis, C.  Schneider, and the other members of
the PEP group for helpful comments.

\appendix
\section{Equivalence of the sum with the convergence factor}\label{prfconv}
Let $p\in\R^3$ and for $k,l\in\Z$ let $\nkl:=(k\lx, l\ly, 0)^T$ and $n=|\nkl|=\sqrt{k^2\lx^2+l^2\ly^2}$.
Then it can be shown that for charge neutral systems
\begin{equation}
  \label{ordern3}
  \frac{1}{|p + \nkl|} + \frac{1}{|p - \nkl|} -
  \frac{2}{n}\,=\,
  \O\left(n^{-3}\right)
\end{equation}
(see for example \cite{deleeuw80b} or \cite{arnold01a}).

Now we rewrite the sum with the convergence factor as
\begin{equation}
  \label{absconvsum}
  \begin{split}
    &\sum_{S=0}^\infty\sum_{k^2+l^2=S}\sump_{j=1}^N
    \frac{q_je^{-\beta|p_j+\nkl|}}{|p_j + \nkl|}\\
    &=\,
    \sum_{k,l\in\Z}\frac{1}{2}\sump_{j=1}^N q_j\left(
      \frac{e^{-\beta|p_j+\nkl|}}{|p_j + \nkl|} +
      \frac{e^{-\beta|p_j-\nkl|}}{|p_j - \nkl|} -
      \frac{2e^{-\beta n}}{n}
    \right)\mpt
  \end{split}
\end{equation}

For any $\alpha_n$ where $\beta|\alpha_n - n|< 1$
\begin{equation}
  \label{expan}
  \frac{e^{-\beta(\alpha_n - n)}}{\alpha_n}
  \,=\, \frac{1}{\alpha_n}
  \sum_{l=0}^\infty \frac{(-1)^l}{l!}\beta^l(\alpha_n - n)^l
  \,=\, \frac{1}{\alpha_n} + \beta\left(\frac{n}{\alpha_n}-1\right) +
  \theta(\beta,\alpha_n,n)
\end{equation}
where $|\theta(\beta,\alpha_n,n)|\le\frac{1}{2}
\frac{\beta^2(\alpha_n - n)^2}{\alpha_n}$.

Therefore for sufficient small $\beta$ we have
\begin{equation*}
  \begin{split}
    &\left|
      \frac{e^{-\beta|p+\nkl|}}{|p+\nkl|} +
      \frac{e^{-\beta|p-\nkl|}}{|p-\nkl|} -
      \frac{2e^{-\beta n}}{n}
    \right|\\
    &\le\, e^{-\beta n}\left|
      \frac{1}{|p+\nkl|} +
      \frac{1}{|p-\nkl|} -
      \frac{2}{n}
    \right|\\
    &\quad + e^{-\beta n}\beta\left|
      \left(\frac{n}{|p+\nkl|} - 1\right) +
      \left(\frac{n}{|p-\nkl|} - 1\right)
    \right|\\
    &\quad + e^{-\beta n}\left|
      \theta(\beta,|p+\nkl|,n) +
      \theta(\beta,|p-\nkl|,n)
    \right|\\
%%%%%%%%%
    &=\, \frac{e^{-\beta n}}{n^3}\O(1)
    + \frac{\beta e^{-\beta n}}{n^2}\O(1)
    + \frac{\beta^2 e^{-\beta n}}{n}\O(1)\mpt
  \end{split}
\end{equation*}
This can be seen by applying equation \eqref{ordern3} to the two first
terms and equation \eqref{expan} to the last.
By approximating the sums over these terms separately as integrals it
is easy to see that the sum in equation \eqref{absconvsum} converges
uniformly for small $\beta$ and therefore
\begin{equation}
  \begin{split}
    &\lim_{\bN}\sum_{k,l\in\Z}\frac{1}{2}\sump_{j=1}^N q_j\left(
      \frac{e^{-\beta|p_j+\nkl|}}{|p_j + \nkl|} +
      \frac{e^{-\beta|p_j-\nkl|}}{|p_j - \nkl|} -
      \frac{2e^{-\beta n}}{n}
    \right)\\
    &=\sum_{k,l\in\Z}\frac{1}{2}\sump_{j=1}^N q_j\left(
      \frac{1}{|p_j + \nkl|} + \frac{1}{|p_j - \nkl|} -
      \frac{2}{n}
    \right)\\
    &= \sum_{S=0}^\infty\sum_{k^2+l^2=S}\sump_{j=1}^N
    \frac{q_j}{|p_j + \nkl|}\mpt
  \end{split}
\end{equation}

\end{document}